%% file: VisionMacroQuant.tex
\documentclass[aps,pra,a4paper, twocolumn, 10pt, nofootinbib]{revtex4-1} % , showpacs
\usepackage{bbm, amsmath, amssymb, amsthm, bm, textcomp,graphicx,color}
\usepackage[utf8]{inputenc}
\usepackage[T1]{fontenc}
\usepackage[english]{babel}
\usepackage{graphicx}

\newcommand{\beq}{\begin{equation}}
\newcommand{\eeq}{\end{equation}}
\newcommand{\beqa}{\begin{eqnarray}}
\newcommand{\eeqa}{\end{eqnarray}}

\newcommand{\ket}[1]{\mbox{$ | #1 \rangle $}}

\def\down{\downarrow}

\def\opone{\leavevmode\hbox{\small1\normalsize\kern-.33em1}}

\begin{document}

%%%% Article title to be placed here
\title{From Quantum Foundations to Applications and Back}

\author{%%%% Author details
Nicolas Gisin, Florian Fr\"owis}

%%%%%%%%% Insert author address here
\affiliation{Group of Applied Physics, University of Geneva, 1211 Geneva 4, Switzerland}

%%%% Abstract text to be placed here %%%%%%%%%%%%
\begin{abstract}
Quantum non-locality has been an extremely fruitful subject of research, leading the scientific revolution towards quantum information science, in particular to device-independent quantum information processing. We argue that time is ripe to work on another basic problem in the foundations of quantum physics, the quantum measurement problem, that should produce good physics both in theoretical, mathematical, experimental and applied physics. We briefly review how quantum non-locality contributed to physics (including some outstanding open problems) and suggest ways in which questions around Macroscopic Quantumness could equally contribute to all aspects of physics.
\end{abstract}
%%%%%%%%%%%%%%%%%%%%%%%%%%%

\maketitle

\section{Introduction}
%===========================
In the 1980's discussing the foundations of quantum theory was truly not fashion. "Bohr solved it all", was the official answer to all questions. Nevertheless, there were some underground activities 
\cite{SundaysQengineer}. And after a few beers, people started confessing their challenges with some quantum concepts. The discussions would mostly concern two problems. First, quantum non-locality, in the sense of violations of Bell inequality: How can these two locations out there in space-time know about each other? Second, the quantum measurement problem: How could an apparatus made out of ordinary matter not obey the superposition principle just because of a sticker saying ``measurement apparatus''?

Despite the negative atmosphere around the foundations, some reckless persons devoted quite some of their time and energy to those two questions, non-locality and the measurement problem. Now, looking backwards almost 40 years, some of those persons got pay back generously --including one of us-- because they were among the very first to understand entanglement (a word one of us never heard during his entire student lifetime!), the no-cloning theorem and further basic concepts of what became a new science: quantum information science.

Today, quantum information is a well established field. Entire journals are devoted to it. The big ones (PRL, Nature, Science) are eager to publish the findings. Every country has some labs devoted to it and universities compete to attract the best players. Even large (and some smaller) companies advertise to hire specialists. The big political blocks invest, and so on. Allow a little satisfaction here: Europe has done very well, at least so far, probably because there was more tolerance for foundations 40 years ago.

It is a fact that the questioning around quantum non-locality has been extremely fruitful. The 1991 Ekert protocol for quantum key distribution \cite{E91} is a heroic example of the breakthroughs that succeeded one after the other in the 1990's, e.g. quantum teleportation in 1993 \cite{Qtelep93} and Shor's celebrated algorithm in 1995 \cite{Shor95}. Could it be that the other problem, the quantum measurement problem, will eventually also prove to be highly fruitful? Is the time ripe?

The answers to these two questions are not obvious, as the two problems are quite different. Nevertheless, we like to argue that the answers are two clear "yes". Moreover, we will argue that the quantum measurement problem is ready to deliver interesting results in all sub-fields of physics, as did quantum non-locality. It will contribute, we believe, to theoretical, mathematical, experimental, applied and conceptual physics.

In the next section we summarize briefly some of the achievements around quantum non-locality, including three outstanding basic questions that remain open. Section \ref{MQ} introduces the quantum measurement problem and underlines how it can contribute to all sub-fields of physics. We argue that the quantum measurement problem might be too loaded with pre-conceptions. Hence, in order to allow for more fresh ideas, one might better talk about ``macroscopic quantumness''. Then, in section \ref{MQexp}, we present a highly subjective review of some experimental results obtained in Geneva in the recent years and a vision for an experimental program.

\section{Quantum Non-Locality: some achievements and 3 outstanding open questions}\label{QNL}
%============================================================================================
The initial question was: Can it really be that Nature requires non-local variables, like, e.g., the quantum state-vector $\Psi$? Schr\"odinger \cite{Schrodinger35} and EPR \cite{EPR35} expressed this question very clearly already in 1935 and Clauser \cite{Clauser72} and Aspect \cite{Aspect82} and co-workers gave the first (and already quite convincing) experimental answers. But physics being an experimental science, a so-called loophole-free experiment was required. Today, we have 4 such experiments \cite{Hanson15,Zeilinger15,Nam15,Weinfurter17}. Some other clever experiments challenged the possibility that some superluminal influences could explain away the non-local quantum correlations \cite{Salart08,Cocciaro11,China13,Cocciaro18}. Such superluminal influences could even be excluded theoretically under the assumption that one can't signal faster than light at the macroscopic level \cite{Bancal12}. Hence, today quantum non-locality is a fact: Nature is able to produce random events that can manifest themselves at several locations, as we like to phrase it. Another (provocative?) sentence is: quantum non-local correlations emerge, somehow, from outside space-time in the sense that no story in space-time can tell how they happen \cite{Salart08,NGScience09,Bancal12}. That's enough for conceptual physics, let's look at the other aspects of non-locality\footnote{We can't resist asking some more provocative questions: How does an event know that it is non-locality correlated to another one? Possibly through the purity of its state? Who keeps track of who is entangled with whom, i.e. where is the information stored? Are there angels manipulating enormously huge Hilbert spaces? Let us emphasize that we take such questions very seriously.}.

Non-locality contributed also to mathematical physics, with the geometrical representation of local correlations by polytopes whose facets represent the Bell inequalities \cite{Pitowski89, RMP-NL-Brunner14}. Theoretical questions about the number of bits that need to be communicated in order to simulate quantum correlations were asked and partly answered \cite{Maudlin92, TappBrassard99,Steiner00, BaconToner03}. But, possibly surprisingly, the most remarkable breakthrough came from applied physics when it was noticed that non-locality, i.e. the certainty that no local variables could describe the observed correlation, guarantees that these correlations contain secrecy, like a cryptographic key as Ekert first noticed \cite{E91}. The intuition is simple and beautiful: if there are no local variables, no one can hold a copy of these non-existing variables. However, true understanding of this took time. For a (long) while it was wrongly understood that quantum key distribution using non-local correlations is equivalent to some prepare-and-measure protocol, like the famous Bennett-Brassard BB84 one \cite{BB84}. But mistakes get corrected and today there is a fascinating sub-field of quantum information dedicated to Device Independent Quantum Information Processing (DIQIP), as it is now called \cite{DIQIPScarani,RMP-NL-Brunner14}. 

Let's turn to the outstanding open questions. We just reminded us that non-local correlations contain secrecy\footnote{Moreover, all non-signaling theories predicting the violation of some Bell inequality have a no-cloning theorem \cite{Masanes06}.}. Indeed, to distribute such correlations one needs shared secret randomness (it can not be generated by local operation and public communication) \cite{Masanes06}, as witnessed by a strictly positive intrinsic information \cite{MW99}:
\beq
I(A:B\down E)\equiv \min_{E\rightarrow F} I(A:B|F)
\eeq
where the minimum is taken over all channels $E\rightarrow F$ from Eve's information $E$ to $F$ and $I(A:B|F)$ denotes the Alice-Bob mutual information conditioned on $F$.
But, more interesting is to extract or distil a secret key out of the non-local correlation. Surprisingly, this can be done using known tools only if the violation of the Bell inequality is large enough. Is it that better distillation tools are waiting for us to discover them? Or is it that there is here a classical analog of bound entanglement: some correlations (between classical bits) require secrecy to be distributed, but this secrecy can never be distilled out of the correlation? That would be some bound-information \cite{GWboundInfo99,GWboundInfo00,GWboundInfo02}. We believe that this is a very difficult question, but a fascinating one, Figure 1.

\begin{figure}
\centerline{\includegraphics[trim=120 100 130 70,clip,width=\columnwidth]{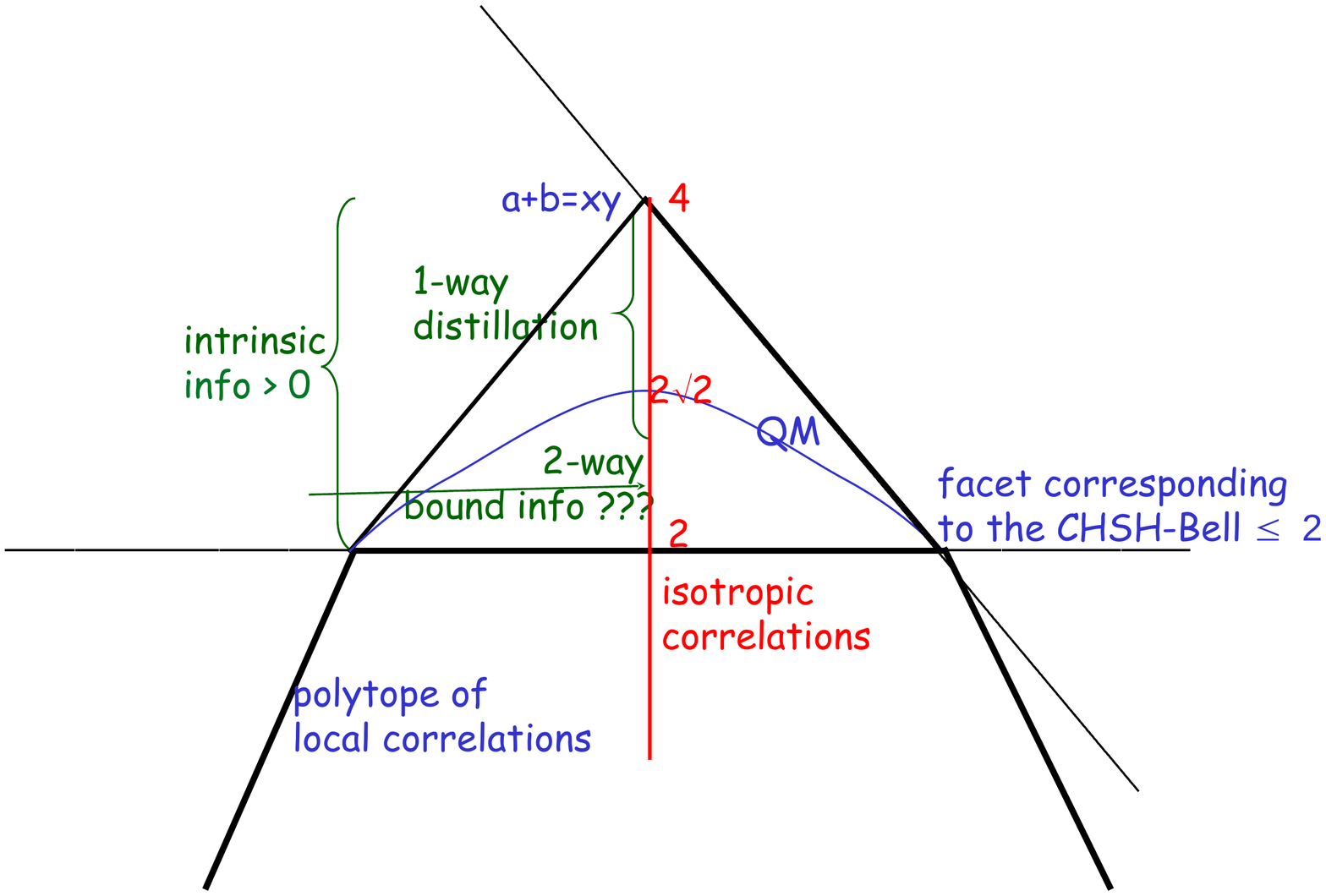}}
\caption{\it Geometrical illustrating of the question of the existence of bound information in low violations of Bell inequality.}
\end{figure}

A second open question concerns just two qubits in a pure state, each undergoing some projective measurement, Figure 2. Can the resulting correlation $p(a,b|\vec x,\vec y)$ be simulated using shared randomness and just one bit of communication? If the two qubits are entangled, Bell tells us that it is impossible to simulate the correlation with no communication. Toner and Bacon showed how this can be done with 2 bits of communication (one bit suffices if the state is maximally entangled) \cite{BaconToner03}. In brief, just one pure partially entangled state of 2 qubits, projective measurements and one classical bit. A simple looking question waiting for an answer.

\begin{figure}
\centerline{\includegraphics[trim=90 350 220 60,clip,width=\columnwidth]{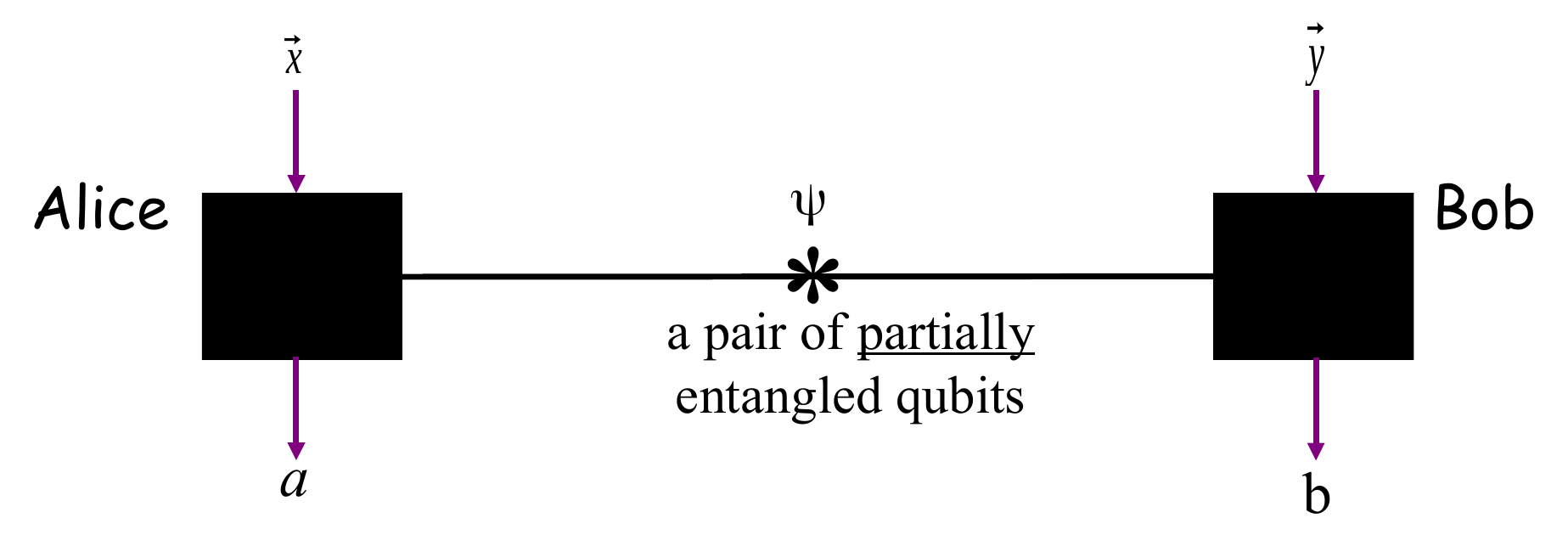}}
\caption{\it An seemingly simple problem: Can one simulate the correlations produced by projective measurements on a 2-qubit partially entangled pure state?}
\end{figure}

Finally, here is a third question that we like very much because it was provoked by applied physics. In applied quantum communication we work hard to develop quantum networks where many nodes would be connected to several neighbors, see Figure 3-left. 

\begin{figure}
  \centerline{\includegraphics[trim=50 150 140 50,clip,width=\columnwidth]{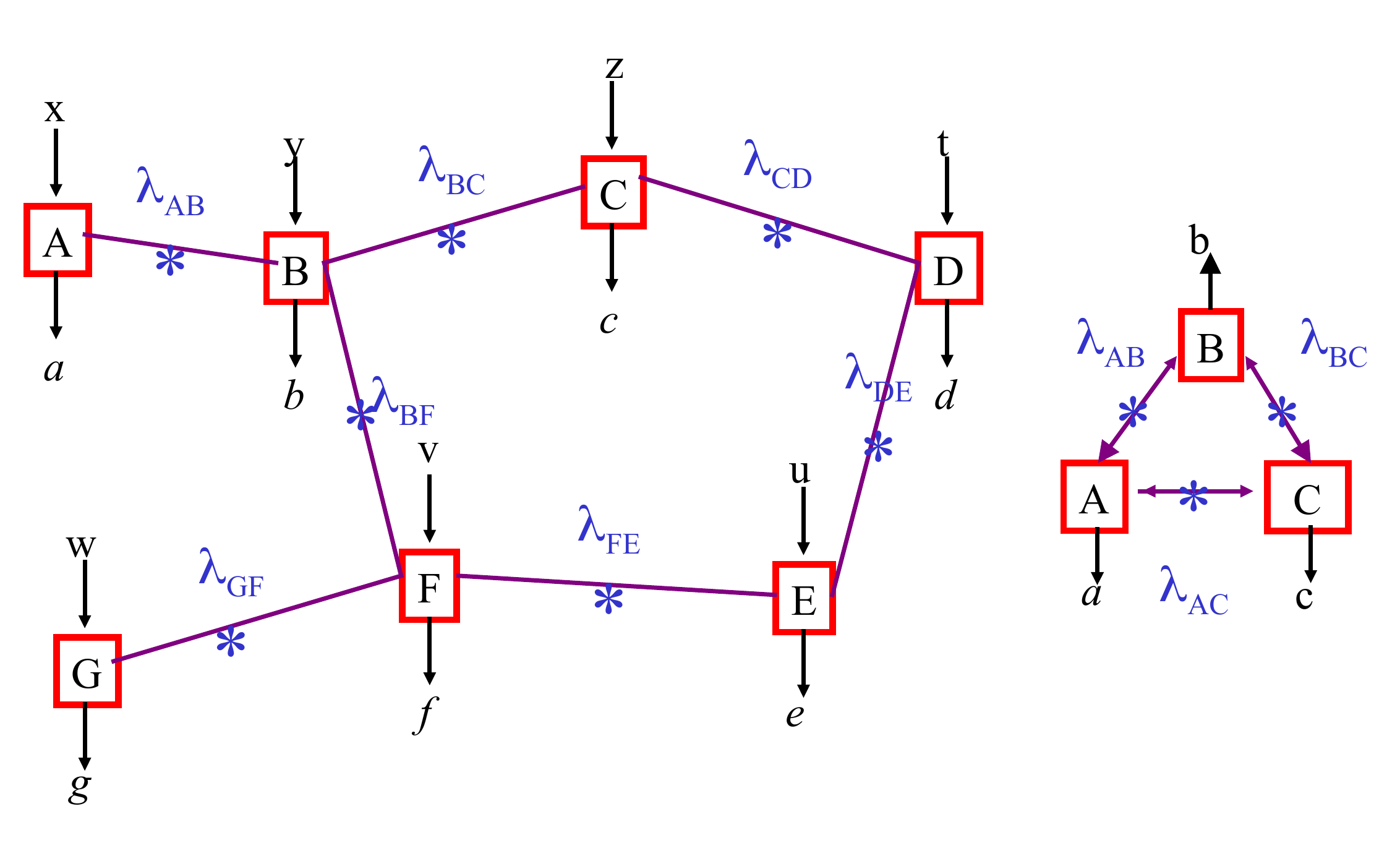}}
\caption{\it Left: example of a quantum network, the $\lambda_{ij}$ could represent a quantum state or local variables. Right: the triangle, the simplest network with a loop, without inputs.}
\end{figure}

After distributing entanglement, each node would undergo some Bell State Measurement in order to extend entanglement to longer and longer distances. Replace each quantum source by a source of local variables, as was done initially by Bell in the case of two parties \cite{Bell64}. Assume that the sources of local variables are independent, similarly to the quantum states in quantum networks that are independent. Which correlations can be simulated by such a local model? What is the equivalence to the standard Bell inequality? A little is known, in particular that the set of local correlations in networks is not convex \cite{biloc10}, hence that the Bell-equivalent inequalities are not linear. For linear chains we have some preliminary result \cite{nlocRosset16}, but not much. In particular no example was found in which the resistance to noise per quantum state is larger in networks than in simple 2-party scenarios. Moreover, in case of (realistic) networks with loops essentially nothing is known, not even for a mere triangle\footnote{except for Fritz's nice example, though this example is essentially the well-known 2-party CHSH scenario forming the base of the triangle with the third node playing the role of the referee who provides the inputs \cite{Fritz12}.} \cite{triangle12,NGtriangle17}, see Figure 3-right. Annoying! Somehow, we feel we still don't understand non-locality.

\section{The Quantum Measurement Problem and Macroscopic Quantumness}\label{MQ}
%==============================================================================
If you believe in reductionism, the quantum measurement problem is easy to formulate: which configuration of atoms and photons constitute a measurement apparatus? Indeed, apparatuses are made out of ordinary atoms, photons and possibly other quantum particles. Hence, either there are some ``magical'' configurations for which the superposition principle stops applying. Or the superposition principle holds for ever and we live in some many-worlds. 

The reasons against many-worlds are explained in \cite{CollapseWhatElse} and against Bohmian mechanics in \cite{CollapseWhatElse,WhyBohmian}, see also why Bohm himself didn't like Bohmian mechanics \cite{Bohm61}. Accordingly, why not give up reductionism? Why not indeed, but with what shall we replace it? The new interpretation Context-System-Modality \cite{CSM18}? Downwards causation, whatever this means precisely \cite{Ellis}? When under strong pressure, one of us declares himself a spontaneous collapse guy. But frankly, we don't know. Nevertheless, we believe that the quantum measurement problem will lead to fascinating new physics, both theoretical, mathematical, experimental and applied physics. Working on all these aspects we'll make progress also on the conceptual problem, as we did with quantum non-locality.

The experimental physics grand goal is clear: demonstrate superpositions of macroscopic objects. But, in some sense, that has already been done. For example one routinely superposes and entangles entire spools of optical fibers. True, each spool of fiber contains only about one excitation, i.e. one photon. Gas cells \cite{Polzik04}, optical crystals \cite{Usmani12}, superconducting qubits \cite{Martinis09}, among others, have been shown to satisfy the superposition principle in ways similar to the optical fiber spools; see also, large molecules \cite{Arndt99}, atoms and photons in a cavity \cite{Haroche01} and nano-mechanical oscillators \cite{nanoOscillators14}. Hence, before we can work on the grand goal, we need to understand what ``macroscopic'' means (see, e.g., our review \cite{macroQRMP17} and the many references therein). More precisely, we need to understand what it means for an object to be macroscopic and quantum in meaningful ways. This is the grand goal for theoretical physics. We don't expect a single good measure, macroscopic quantumness has many facets, indeed. But there is a need for structuring these factes and combining the macro side with the quantum side in relevant figures of merit. There is already some literature on this, but quite disconnected from experiments. There are also some experiments, as those mentioned above and quite more, but usually disconnected from the abstract theoretical concepts. Technology is ready, we can now work in good synergy between experiments and theory towards the grand goals. This is the main message of this paper.

Mathematical physics has also a role to play, probably in providing useful tools to quantify the relation between the various measures of macroscopic quantumness and decoherence. Another goal is to improve our understanding of non-Markovian evolutions of open quantum systems. Indeed, the better we isolate large quantum systems, the more likely it is that the remaining interactions with the environment lead to non-Markovian effects.

For applied physics our guess is that quantum sensing will be the main application. Large quantum systems are notoriously fragile, hence, with appropriate control, should provide particular sensitivity to all kinds of physical quantities.

Finally, we like to suggest to use the terminology of Macroscopic Quantumness. This captures what we need to work on today. It is less loaded with all sorts of historical prejudices than when we talk about the measurement problem (like people working on Device-Independent Quantum Information Processing don't need to care about the futile debate on the terminology "non-locality"). And those who care about the measurement problem can use the terminology of Macroscopic Quantumness without stress.

\section{Examples of experimental Macroscopic Quantumness}\label{MQexp}
%======================================================================
Nowadays, two entangled qubits is a very well understood concept (though, see the 2nd open question in section 2) and is widely mastered in many experimental situations. Hence, let's make the qubits bigger in one out of many possible ways, see Figure 4. Below is a highly subjective and partial list of some of our achievements, mostly during last year:
\begin{enumerate}
\item One can replace one of the qubit by a bigger object. In \cite{MacroMicro12} we displaced a single-photon up to 500 photons, hence demonstrate entanglement between 500 photons on one side with a single-photon on the other side (see also the related experiment by the Calgary group \cite{MacroMicrocalgary}). We could also realize a similar experiment in our quantum memories, though limited to about 30 photons \cite{MacroMicro16}.
\item One could increase the Hilbert space dimension of each sub-system. In \cite{highDim17} we used trains of time-bins and energy-time entanglement to demonstrate up to more than 4 e-bits (entanglement-bits), not a huge number but to the best of our knowledge the highest value. Clearly, there is room for improvements. Further, in \cite{highDimQM17} we demonstrated more than 2 e-bits in a solid-state quantum memory. Again, there seems to be plenty of space for improvements.
\item One may increase the number of parties and look for large entanglement depth (i.e. large numbers of parties within one entangled state). For example, in \cite{certifMillionEnt17} data from an experiment using our solid-state quantum memory allowed us to certify an entanglement depth of more than 16 million ions.
\end{enumerate}
And much more is possible, there is no limit but imagination. The goal of the above list is mostly to encourage more people to enter the field. It is in no way a review of the fascinating and highly varied activities going on around the globe. Just a little selfish summary.

\begin{figure}
\centerline{\includegraphics[width=\columnwidth]{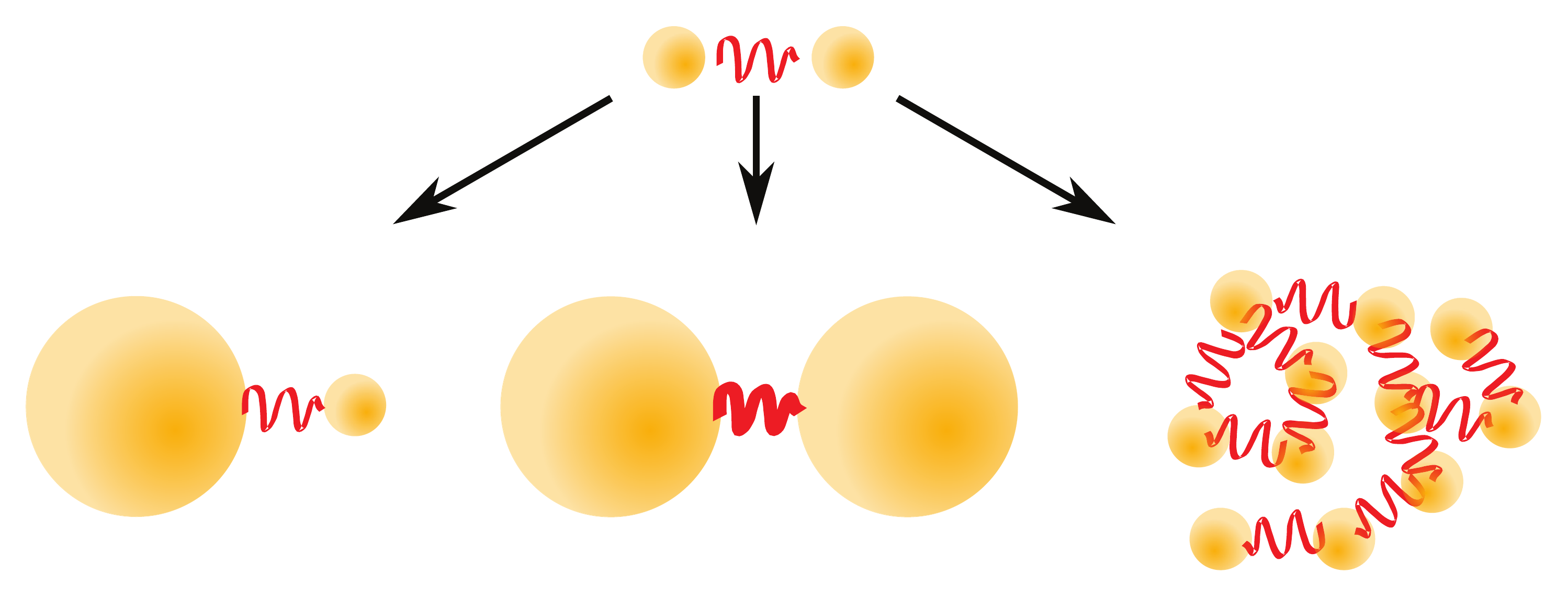}}
\vspace{-10pt}
\caption{\it Illustration of several ways in which the common 2-qubit entanglement can be made big. From left to right: make one side hold many excitations, make the Hilbert space of each party large, increase the number of parties involved in one entangled "block" (i.e. increase the entanglement depth).}
\end{figure}

A possible vision is to combine all of the above, i.e. increase simultaneously the number of entangled ions, the number of excitations and the Hilbert space dimension. For example, Figure 5 illustrated a train of single-photons, each first hitting a beam-splitter, hence producing a train of entangled optical modes. Next, each mode is displaced by some hundreds of photons\footnote{thanks to a corresponding train of coherent states $\ket{\sqrt{2}\alpha}$ entering the second input port of the beam splitter.} and then stored in some solid-state quantum memories. At this point one would have hundreds of e-bits, thousands of excitations, an entanglement depth of millions and billions of involved ions. Admittedly, these are not yet truly macroscopic numbers, but there seems to be a clear path towards "large" numbers, i.e. towards macroscopic quantumness.

\begin{figure}
\centerline{\includegraphics[trim=70 250 200 90,clip,width=\columnwidth]{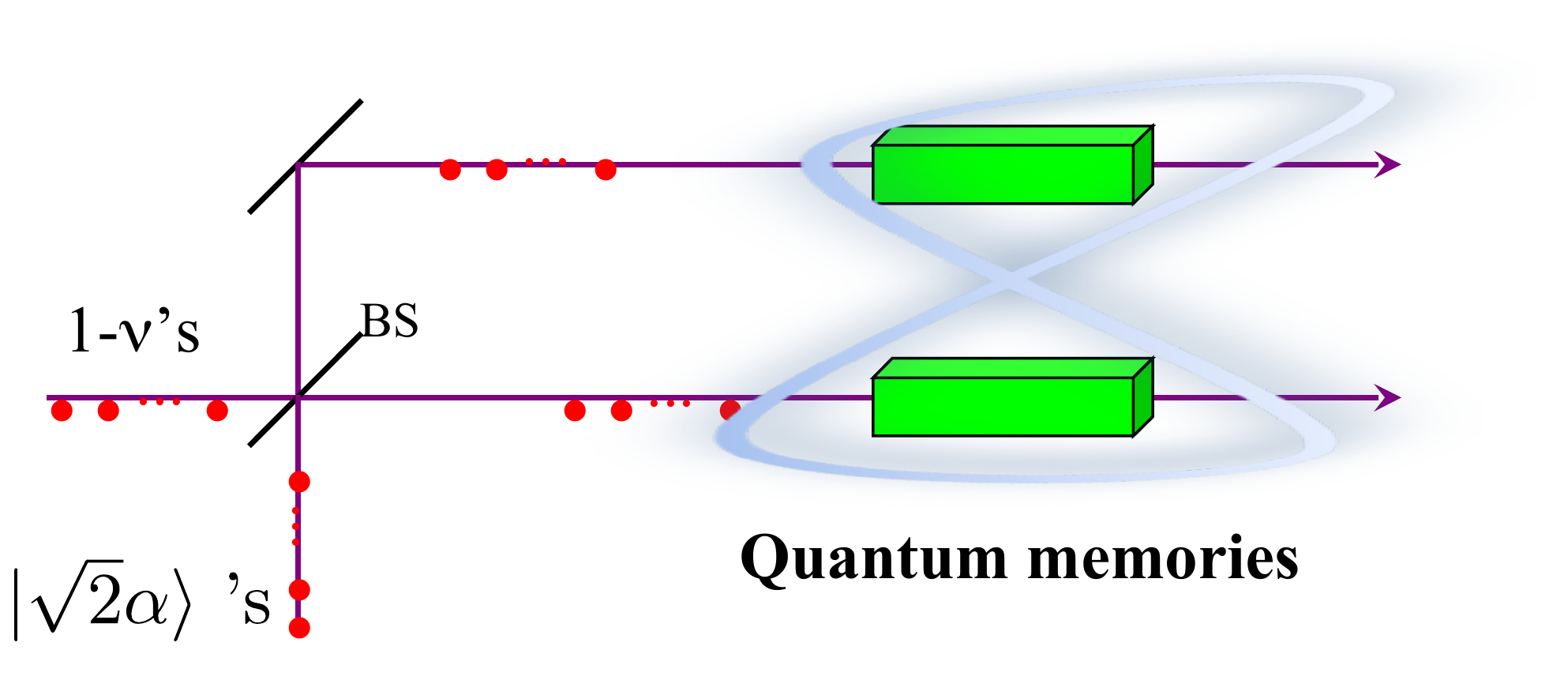}}
\caption{\it Illustration of possibility to demonstrate Macroscopic Quantumness with hundreds of e-bits, thousands of excitations, an entanglement depth of millions and billions of ions, see text.}
\end{figure}

\section{Conclusion}
%===================
Technology is ripe to demonstrate macroscopic quantumness and guide theory towards meaningful figures of merit. Also mathematical and applied physics will contribute to this timely research field. In the background there is the quantum measurement problem, a bit like quantum non-locality inspired quantum information science. After briefly reviewing the odd situation of quantum foundations up to the 1980's and the revolution in the 1990's (including 3 open problems that we consider as especially outstanding), we argue that "the other foundational problem", namely the measurement problem, holds the potential for inspiring a revolution of similar amplitude. We illustrated briefly some possible vision, though the ongoing research in macroscopic quantumness is already much broader.

\section*{Acknowledgements}
Financial support by the European ERC-AG MEC and the Swiss NSF are gratefully acknowledged.

%%%%%%%%%% Insert bibliography here %%%%%%%%%%%%%%

\input{biblioRSTA_Gisin_tex.tex}

\end{document}

%% file: biblioRSTA_Gisin_tex.tex
%%% Local Variables:
%%% mode: latex
%%% TeX-master: "RSTA_Gisin_tex"
%%% End: